\documentclass[onecolumn,showpacs,preprint,amsmath,amssymb,superscriptaddress]{revtex4-1}

\usepackage{dcolumn}% Align table columns on decimal point
\usepackage{bm}% bold math

\newcommand{\ignore}[1]{}
\usepackage{graphicx}
\usepackage{amsmath,amssymb,amsfonts,amsthm,color,latexsym}
\usepackage{soul}

\def\mb{\mathbf}

\def\bm{\boldsymbol}
\newcommand{\Rmnum}[1]{\uppercase\expandafter{\romannumeral #1\relax}}

\begin{document}

\preprint{}

\title{Data-driven parameterization of the generalized Langevin equation}% Force line breaks with \\

\author{Huan Lei}
\email{huan.lei@pnnl.gov}
\affiliation{Pacific Northwest National Laboratory, Richland, WA 99352.}%
\author{Nathan Baker}
 \email{Nathan.Baker@pnnl.gov}
\affiliation{Pacific Northwest National Laboratory, Richland, WA 99352.}%
\affiliation{Division of Applied Mathematics, Brown University, Providence, RI 02912.}%
\author{Xiantao Li}
 \email{xli@math.psu.edu}
\affiliation{The Pennsylvania State University, University Park, PA 16802.}%


\begin{abstract}
We present a data-driven approach to determine the memory kernel and random noise in generalized Langevin equations. 
To facilitate practical implementations, we parameterize the kernel function in the Laplace domain by a rational function, with coefficients directly linked to the equilibrium statistics of the coarse-grain variables.
We show that such an approximation can be constructed to arbitrarily high order and the resulting generalized Langevin dynamics can be embedded in an extended stochastic model without explicit memory.
We demonstrate how to introduce the 
stochastic noise so that the second fluctuation-dissipation theorem is exactly satisfied. Results from several numerical tests are presented to demonstrate the effectiveness of the proposed method.
\end{abstract}

% check this

\maketitle
\section{Introduction}

Generalized Langevin equations (GLEs) have recently re-emerged in the area of molecular modeling as a promising description of reduced-dimension coarse-grained variables.
In principle, GLEs can be derived using the Mori-Zwanzig projection formalism \cite{Mori65, Zwanzig73}.
Examples of such derivations can be found for a variety of applications \cite{berne1990dynamic, ChSt05, curtarolo_dynamics_2002, hijon2010mori, IzVo06, Li2009c, MaLiLiu16, givon2004extracting}; e.g., climate modeling \cite{wouters2013multi, majda2012physics}.
However, practical implementations of GLEs require specification of the memory function which can be difficult to obtain, even when the full dynamics of the system is known.
For example, the memory functions obtained in past studies \cite{Li2009c, MaLiLiu16, LiXian2014, Darve_PNAS_2009} have involved functions of high-dimensional matrices.
Darve \textit{et al.} proposed a more efficient algorithm \cite{Darve_PNAS_2009} to compute the memory kernel by solving an equation for the orthogonal dynamics derived from the Mori-Zwanzig formalism.
However, the orthogonal dynamics equation can be expensive to solve when the original system is large.
Furthermore, even when the memory kernel function is available, direct evaluation of the memory term can be costly because it requires the history of the coarse-grained (CG) variables at every time step and the associated numerical integration.
Sampling of the random noise is also a challenging component of GLEs: to generate the correct equilibrium statistics for the CG model, the random noise has to obey the second fluctuation-dissipation theorem (FDT) \cite{Kubo66}.
The theory of stationary processes  \cite{Doob44} states that the random process is uniquely determined by the correlation function, which is proportional to the memory kernel; however, sampling the random noise is nontrivial in practice.
Methods based on matrix factorization are computationally challenging because they require decomposition of a correlation matrix with dimension proportional to the total simulation period.
Alternatively, more efficient methods based on Fast Fourier Transforms (FFTs) may create artificial periodicity \cite{li2015incorporation}. 

In addition to the direct derivation of memory kernels \cite{Li2009c, MaLiLiu16, LiXian2014, Darve_PNAS_2009}, there have been numerous attempts to compute the memory kernel from full molecular dynamics (MD) simulations \cite{oliva2000generalized, berkowitz1981memory, berkowitz1983generalized}, especially for systems with zero net mean force.
Such analyses lead to integral equations of the first kind, which are numerically unstable without additional regularization.
% More importantly, it still faces the issues with the random noise and the evaluation of the memory term.
Another approach for estimating the kernel uses Kalman filtering and assumes functions of exponential form \cite{harlim2015parametric,fricks2009time} such that the GLE can be embedded in a Markovian dynamics framework.
In recent work, Chorin and Lu \cite{chorin2015discrete} considered a time-discrete representation, representing the memory effects using the NARMAX (nonlinear autoregression moving average with exogenous input) method.
Voth \textit{et al.} proposed an alternative approach \cite{Dav_Voth_JCP_2015} to recover the coarse-grained dynamics by introducing fictitious particles that interact with the coarse-grained variables, effectively introducing an approximation of the kernel function. 

In this work, we present a hierarchical approach to obtain GLE kernel functions from simulation data. 
Such a data-driven approach is particularly useful for complex models (e.g., biomolecules or climate) in which full dynamics models are typically unavailable or inaccessible with finite computing resources.
The key idea is parameterization of the kernel function Laplace transform by a rational approximation.
The first two approximations in the rational approximation hierarchy correspond to the Markovian approximation \cite{hijon2010mori,Lei_Cas_2010,kauzlaric2011bottom} and approximation of noise by a Ornstein-Uhlenbeck  process, which is the ansatz used in some previous works.
As we will show, these two approximations are often insufficient to predict dynamics properties; however, our hierarchy can be used to construct arbitrarily high-order models.  
The parameters in our ansatz can be computed from statistical properties of the coarse-grained variables.
Additionally, this ansatz makes it possible to eliminate memory from the GLE by introducing auxiliary variables.
In particular, we will show how to introduce inexpensive white noise terms into the extended dynamics to approximate the random noise while satisfying the second FDT is {\it exactly}.
As a result, no memory term needs to be evaluated, and no colored noise needs to be sampled.

\section{Numerical Method}

The generalized Langevin equation (GLE) can be expressed in the following form
\begin{equation}\label{eq: gle}
	\left\{
	\begin{aligned}
		\dot{\mb{q}}&= M^{-1}\mb{p},\\
		\dot{\mb{p}}&= \mb{F}(\mb{q}) - \int_0^t \bm{\theta}(t-\tau) \mb{v}(\tau)  d\tau + \mb{R}(t),
	\end{aligned}\right.
\end{equation}
where $\mb{q}$ and $\mb{p}$ are the coarse-grained coordinates and momenta, and $\mb{F}(\mb{q}) = -\nabla U(\mb{q})$ is the conservative force term determined by the potential of mean force (PMF) $U(\mb{q})$.
$\bm{\theta}(t)$ denotes a memory kernel function, which is the main focus of this paper.
The noise $\mb{R}(t)$ is a stationary Gaussian process with zero mean, satisfying the second fluctuation-dissipation theorem (FDT) \cite{Kubo66}:
\begin{equation}\label{eq: fdt}
	\langle \mb{R}(t) \mb{R}(t')^T \rangle= \beta^{-1} \bm{\theta}(t-t'),
\end{equation}
where $\beta = 1/k_BT$. 
% In Eq. \eqref{eq: gle}, $\mb{q}$ and $\mb{v} = \mb{p}/m$ could come from a coarse-graining process, in which case, they represent the coarse-grained variables. 
Such equations have been derived in previous work \cite{Li2009c, LiXian2014} using the Mori-Zwanzig formalism \cite{Mori65, Zwanzig61}.
In the present work,
% we assume the form of GLE, and instead
our main goal is to estimate the memory kernel and noise terms from dynamics simulation data.
% We will choose an appropriate parametric form for the kernel function to make the implementation more straightforward.

We assume that we have a time series data set of $\mb{v}$ and $\mb{F}(\mb{q})$ ($\mb{v}=M^{-1}\mb{p}$) drawn from an equilibrium simulation such that the time series corresponds to a stationary random process. 
We right-multiply the second equation in \eqref{eq: gle} by $\mb{v}(0)^T$ to obtain
\begin{equation}\label{eq: gh}
	\mb{g}(t)= \int_0^t \bm{\theta}(t-\tau) \mb{h}(\tau) d\tau.
\end{equation}
Here we have defined the correlation matrices,
\begin{equation}
	\mb{g}(t)= \langle M \dot{\mb{v}}(t) - \mb{F}(\mb{q}(t)),  \mb{v}(0)^T\rangle, \mb{h}(t)= \langle \mb{v}(t),  \mb{v}(0)^T\rangle, 
\end{equation}
and we have made the assumption that $\langle \mb{R}(t) \mb{v}(0)^T \rangle=0$, which was verified in previous work \cite{LiXian2014}. 

Given the correlation functions, Eq.~\eqref{eq: gh} can be regarded as an integral equation from which the memory function can be computed.
However, this is an integral equation of the first kind, and it is not well-posed, leading to unreliable solutions.
Instead of a determining the kernel function directly in the time-domain, we can instead parameterize its Laplace transform.
% As we will show, this approach offers several advantages. 
% , in which we construct the memory kernel through Laplace transform, which is equivalent to the Krylov subspace approximation that was developed in our previous work \cite{LiXian2014}.  
We define the Laplace transform as
\begin{equation}\label{eq: Lap}
	\mb{G}(\lambda)= \int_0^{+\infty} \mb{g}(t) e^{-t/\lambda} dt.
\end{equation}
Similarly, we denote the Laplace transforms of $\mb{h}(t)$ and $\bm{\theta}(t)$ by $\mb{H}(\lambda)$ and $\mb{\Theta}(\lambda)$, respectively.
Taking the Laplace transform of Eq.~\eqref{eq: gh}, we arrive at
\begin{equation}
	\label{eq:theta_lambda}
	\mb{G}(\lambda) = -\bm{\Theta}(\lambda) \mb{H}(\lambda).
\end{equation}
%\st{The idea is to first determine the value of $\bm{\Theta}$ at zero and infinity.}
We utilize the values of $\bm{\Theta}$ at specific time points to construct $\bm{\Theta}(\lambda)$.
By taking $\lambda \to +\infty$, we obtain 
%$\bm{\Theta}(+\infty)= -\mb{G}(+\infty) \mb{H}(+\infty)^{-1}.$
\begin{equation}\label{eq: th-inf}
	\bm{\Theta}(+\infty)= -\mb{G}(+\infty) \mb{H}(+\infty)^{-1}.
\end{equation}
It is clear that
\begin{equation}
	\mb{G}(+\infty)=\int_0^{+\infty} \mb{g}(t) dt, \quad \mb{H}(+\infty)=\int_0^{+\infty} \mb{h}(t) dt
\end{equation}
which makes it possible to incorporate a Green-Kubo type of formula in the approximation and model accuracy over long time scales.

For short or intermediate time scales, we use the point $\lambda = 0$. 
Using \eqref{eq:theta_lambda}, one can find the limiting values of the kernel and its derivatives as $\lambda \to 0$. 
%Direct calculations yield,
%\begin{equation}
%\bm{\Theta}(0)= -\mb{G}'(0) \mb{H}'(0)^{-1}. 
%\end{equation}
% and values at a series of discretized (or selected?) points. 
This calculation amounts to computing $\mb{G}'(0),$ and similarly $\mb{H}'(0)$, which is straightforward.
For instance, by integrating by parts repeatedly in \eqref{eq: Lap}, we find that
\begin{equation}
	\mb{G}'(0)=\mb{g}(0), \frac{\mb{G}''(0)}2=\mb{g}'(0),  \cdots, \frac{\mb{G}^{(j)}(0)}{ j!}= \mb{g}^{(j-1)}(0).
\end{equation}
For example, we have $\mb{H}'(0)=\mb{h}(0)=k_B T \mb{I}$.
In addition, we can find $\bm{\Theta}(0)=0,$ 
%$\bm{\Theta}'(0) = -\beta \mb{g}'(0)$, and
\begin{equation}\label{eq: th-der}
	\begin{split}
		%\bm{\Theta}'(0)= \frac12 \beta \mb{g}'(0),
		\bm{\Theta}'(0) &= -\beta \mb{g}'(0), \\
		\bm{\Theta}''(0) &= -2 \beta \big[\mb{g}''(0)+\beta \mb{g}(0) \mb{h}''(0)\big], \\
		\bm{\Theta}'''(0) &= -6 \beta \big[\mb{g}'''(0)+\beta \mb{g}'(0) \mb{h}''(0)\big].
	\end{split} 
\end{equation}
%and
%\begin{equation}\label{eq: th-zero''}
% %\bm{\Theta}''(0)= \frac13 \beta \mb{g}'(0)\big[\mb{g}''(0)-\beta \mb{g}(0) \mb{h}''(0)\big]
% \bm{\Theta}''(0)= 2 \beta \big[\mb{g}''(0)-\beta \mb{g}(0) \mb{h}''(0)\big]
%\end{equation}
%{\color{blue} We may have to just state the result:}
In the derivations above, we have incorporated the values at $\lambda=0$ and $\lambda=+\infty.$
But the ansatz of the rational approximation is quite flexible, and other interpolation points can be used as well.
For stationary process of Hamiltonian system,  $\langle \mb{v}(0) \mb{q}(0)^T\rangle = 0$; therefore, $\mb{g}''(0) = \langle m\ddot{\mb{v}}(0) - \dot{\mb{F}}(\mb{q}(0)), \dot{\mb{v}}(0)^T\rangle \equiv 0$ and $\bm{\Theta}''(0) \equiv 0$.

Given limiting values available extracted from the data, 
%To construct $\bm{\Theta}(\lambda)$, 
%\st{A very robust approximation for these problems is the rational approximation}, 
we seek a rational function approximation for $\bm{\Theta}(\lambda)$,
% that interpolates $\bm{\Theta}(\lambda)$ with the interpolation 
%conditions Eqs. \eqref{eq:theta_lambda}, \eqref{eq: th-inf} and \eqref{eq: th-der}.
in the form of
\begin{equation}
	\begin{aligned}
		\bm{\Theta}(\lambda) \approx&  \big[I - \lambda \mb{B}_0 - \lambda^2 \mb{B}_1 - \cdots -\lambda^n \mb{B}_{n-1}\big]^{-1} \\
		&\times \big[\mb{A}_0 + \lambda \mb{A}_1 + \cdots +  \lambda^{n-1} \mb{A}_{n-1} \big] \lambda.
	\end{aligned}
	\label{eq:Theta_second}
\end{equation}
The coefficients $\{\mb{A}_i,\mb{B}_i\}$ can be determined by matching the limits of $\bm{\Theta}(\lambda)$.
The matching conditions lead to a linear system of equations, which can be solved analytically for small $n$ or numerically for large $n$.

The zero-order approximation treats $\bm{\Theta}(\lambda) \equiv \bm{\theta}_0$ as a constant matrix set to $\bm{\Theta}(+\infty)$. 
Accordingly, one gets a Markovian approximation by a Langevin dynamics with damping coefficient given by $\gamma=\bm{\theta}_0$.
%$\bm{\theta}(t)$ takes the Markovian approximation $\bm{\theta}(t) = \bm{\theta}_0\delta(t)$. 
We can determine $\bm{\theta}_0$ by Eq.~\eqref{eq: th-inf}.
In fact, $\mb{H}(+\infty)$ is proportional to the diffusion tensor; i.e., the matching condition recovers the Einstein relation $D = \displaystyle \frac{k_BT}{\gamma}$
\cite{Kubo66, Einstein_1905}.
% where $\gamma$ is the friction coefficient of the resulting Langevin dynamics.
%$D = \frac{k_BT}{\gamma}$ for a brownian particle, where $D$ and $\gamma$ are the diffusion and
%friction coefficient, respectively.

For the first order approximation ($n=1$), we have
\begin{equation}\label{eq: eg1}
	\bm{\Theta}(\lambda)= \big[I - \lambda \mb{B}_0\big]^{-1} \mb{A}_0 \lambda.
\end{equation}
%By matching \eqref{eq: th-der}
%By matching $\bm{\theta}(+\infty)$ and $\bm{\Theta}'(0)$, we find that,
By matching Eq.~\eqref{eq: th-inf} and Eq. \eqref{eq: th-der}, we find that
\[ \mb{A}_0= \bm{\theta}(0), \quad \mb{B}_0=  -\bm{\theta}(0) \bm{\Theta}(+\infty)^{-1}.\]
In this case, the memory function in the time domain is given by,
\begin{equation}\label{eq: 1st}
	\bm{\theta}(t)\approx e^{t \mb{B}_0} \mb{A}_0. 
\end{equation}
Depending on the eigenvalues of $\mb{B}_0$, the memory function can exhibit both exponential decay and oscillations. 

A computational difficulty in GLE simulation is that the integral has to be evaluated at every step. 
However, this difficulty can be removed by introducing auxiliary equations based on the rational approximation of the memory function \eqref{eq:Theta_second}.
More specifically, we can define
\( \mb{d}(t) = \displaystyle \int_0^t \bm{\theta}(t-\tau) \mb{v}(\tau) d\tau.\)
Then, Eq.~\eqref{eq: eg1} implies that the approximate GLE can be written as,
\begin{equation}\label{eq: gle1}\left\{
	\begin{split}
		\dot{\mb{q}}&= M^{-1}\mb{p}, \\
		\dot{\mb{p}}&= \mb{F}(\mb{q}) + \mb{d}, \\
		\dot{\mb{d}}&= \mb{B}_0\mb{d} -\mb{A}_0\mb{v} + \mb{W}(t),
	\end{split}\right.
\end{equation}
where we added a white noise term $\mb{W}(t)$ satisfying $\langle \mb{W}(t) \mb{W}(t')^T\rangle= -\beta^{-1} (\mb{B}_0\mb{A}_0 + \mb{A}_0 \mb{B}_0^T)
\delta(t-t')$. 
We pick the initial state of the auxiliary variable $\mb{d}$ to satisfy $\langle \mb{d}(0) \mb{d}(0)^T\rangle = \beta^{-1} \mb{A}_0$.

We can show that this new memory-less dynamics corresponds to an approximation of the GLE \eqref{eq: gle}.
The memory function is approximated by the rational function in the frequency space, which is precisely \eqref{eq: 1st}.
More importantly, with this proper choice of the initial condition for $\mb{d}$, the random noise $\mb{R}(t)$ is given by $\mb{R}(t) = \displaystyle \int_0^t e^{\mb{B}_0(t-s)}  \mb{W}(s) ds + \mb{d}(0) e^{\mb{B}_0t},$ which is a stationary Gaussian process that satisfies the second FDT \eqref{eq: fdt} {\it exactly} with an invariant distribution given by
\begin{equation}
	\rho(\mb{q},\mb{p},\mb{d}) \sim e^{-\beta\left(\frac{1}{2}M^{-1}\mb{p}^2 + U(\mb{q}) +\frac{1}{2}\mb{d}^T\mb{A}_0^{-1}\mb{d}\right)}
\end{equation}

The procedure above can be extended to arbitrarily high order, and the extended system can be written as follows,
\begin{equation}\label{eq: gle2}\left\{
	\begin{split} \dot{\mb{q}}&= M^{-1}\mb{p},\\
		\dot{\mb{p}}&= \mb{F}(\mb{q}) + \mb{Z}^T\mb{d},\\
		\dot{\mb{d}} &= \mb{B} \mb{d}  - \mb{Q} \mb{Z} \mb{v} + \mb{W}(t), 
	\end{split}\right.
\end{equation}
where $\mb{B}$ is a matrix and $\mb{W}$ is added white noise.
%, $\mb{Q}$ will be determined later. 
For example, for the fourth-order method, $\mb{B}$ is given by:
\begin{equation}
	\mb{B} = \begin{pmatrix}
		0 & ~0 & ~0 & ~\mb{B}_3 \\
		\mb{I} & ~0 & ~0 & ~\mb{B}_2 \\
		0 & ~\mb{I} & ~0 & ~\mb{B}_1 \\
		0 & ~0 & ~\mb{I} & ~\mb{B}_0
	\end{pmatrix}.
\end{equation}
The matrices $\mb{Q}$ and $\mb{Z}$ can be determined by matching the Laplace transform of the memory kernel given by $\bm{\theta}(t) = \mb{Z}^T e^{\mb{B} t} \mb{Q} \mb{Z}$ with the rational approximation \eqref{eq:Theta_second}.
Similar to Eq.~\eqref{eq: gle1}, we can also show that, by choosing the white noise $\mb{W}$ and the initial conditions of $\mb{d}$ properly; i.e.,
$\langle \mb{d}(0) \mb{d}(0)^T \rangle = \beta^{-1} \mb{Q}$,
$\langle \mb{W}(t)\mb{W}(s)^T \rangle = -\beta^{-1}(\mb{BQ} + \mb{QB}^T)\delta(t-s)$, 
the colored noise generated by the extended dynamics also satisfies the second FDT \eqref{eq: fdt} exactly.
Eq.~\eqref{eq: gle2} has an invariant distribution  given by
\begin{equation}
	\rho(\mb{q},\mb{p},\mb{d}) \sim e^{-\beta\left(\frac{1}{2}M^{-1}\mb{p}^2 + U(\mb{q}) +\frac{1}{2}\mb{d}^T\mb{Q}^{-1}\mb{d}\right)}.
\end{equation}

\section{Numerical Results}

\begin{figure}[htpb]
	\center
	\includegraphics[scale=0.36]{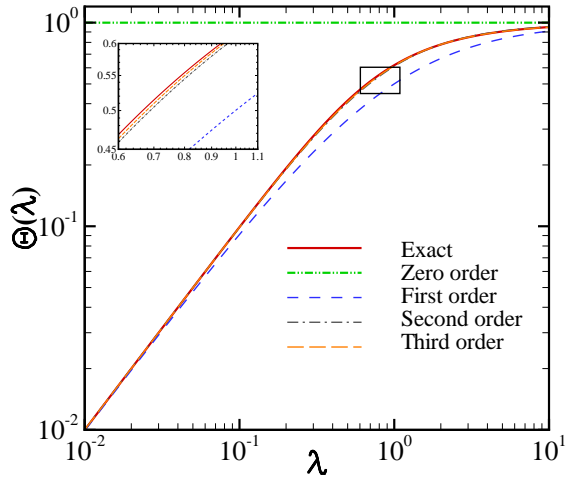}
	\caption{Laplace transform of the memory kernel $\Theta(\lambda)$ modeling a tagged particle attached to a one-dimensional harmonic chain.
	The inset plot shows the closeup view of $\Theta(\lambda)$ within the rectangular region.
	%Lower-right inset shows the velocity correlation functions obtained from GLE with %kernel modeled by different orders of rational approximation.
	}
	\label{fig:1D_chain}
\end{figure}

We demonstrate our method through coarse-graining the dynamics of a tagged particle within a one-dimensional harmonic chain.
In particular, we consider the first particle (on the free end) as the target particle and treat the remaining particles as the heat bath.
As shown in Ref.~\cite{AdDo74,LiE07}, the dynamics of the target particle can be modeled by a GLE with kernel given by 
%\begin{equation}
$\theta(t) = \frac{\sqrt{m}K}{t}J_1\left(\frac{2Kt}{\sqrt{m}}\right)$,
%\end{equation}
where $m$ is the mass of each particle, $K$ is the force constant of harmonic interaction and $J_1$ is a Bessel function of the first kind.
We calculated the trajectory of the tagged particle in a harmonic chain consisting of $N = 1000$ particles with $K$, $m$ and $\beta$ set to unity. Fig.~\ref{fig:1D_chain} shows the numerical results for $\Theta(\lambda)$ obtained using the rational approximation method described above.
As the approximation increases to third order, $\Theta(\lambda)$ agrees well with the exact formulation, which is given by $\frac{\sqrt{1+4\lambda^2}-1}{2\lambda}$.

We also studied a tagged particle immersed in a fluid system governed by pairwise-conservative forces similar to those in dissipative particle dynamics (DPD) simulations \cite{Hoogerbrugge_SMH_1992,Espanol_SMO_1995}; i.e., defined by
\begin{equation}
	\label{eq:DPD_eq2}
	\bm F_{ij} = 
	\begin{cases}
		a(1.0 - r_{ij}/r_c) \bm e_{ij}, & r_{ij} < r_c,\\
	    0, &r_{ij} > r_c,
	\end{cases}
\end{equation}
%-------------------------------------------------------------------------------
where $\bm r_{ij}=\bm r_i-\bm r_j$, $r_{ij} =|\bm r_{ij}|$ and $\bm e_{ij} = \bm r_{ij}/r_{ij}$ and $a$ is the force magnitude and $r_c$ is the cut-off radius beyond which all interactions vanish.
One tagged particle and $2999$ solvent particles were placed in a cubic box $10\times10\times10 r_c^3$ , with a periodic boundary condition imposed in each direction.
The mass $m$ of both the tagged particle and solvent particle were chosen to be unity.
A Nos\'{e}-Hoover thermostat was used to equilibrate the system.

\begin{figure}[htpb]
	\center
	\includegraphics[scale=0.25]{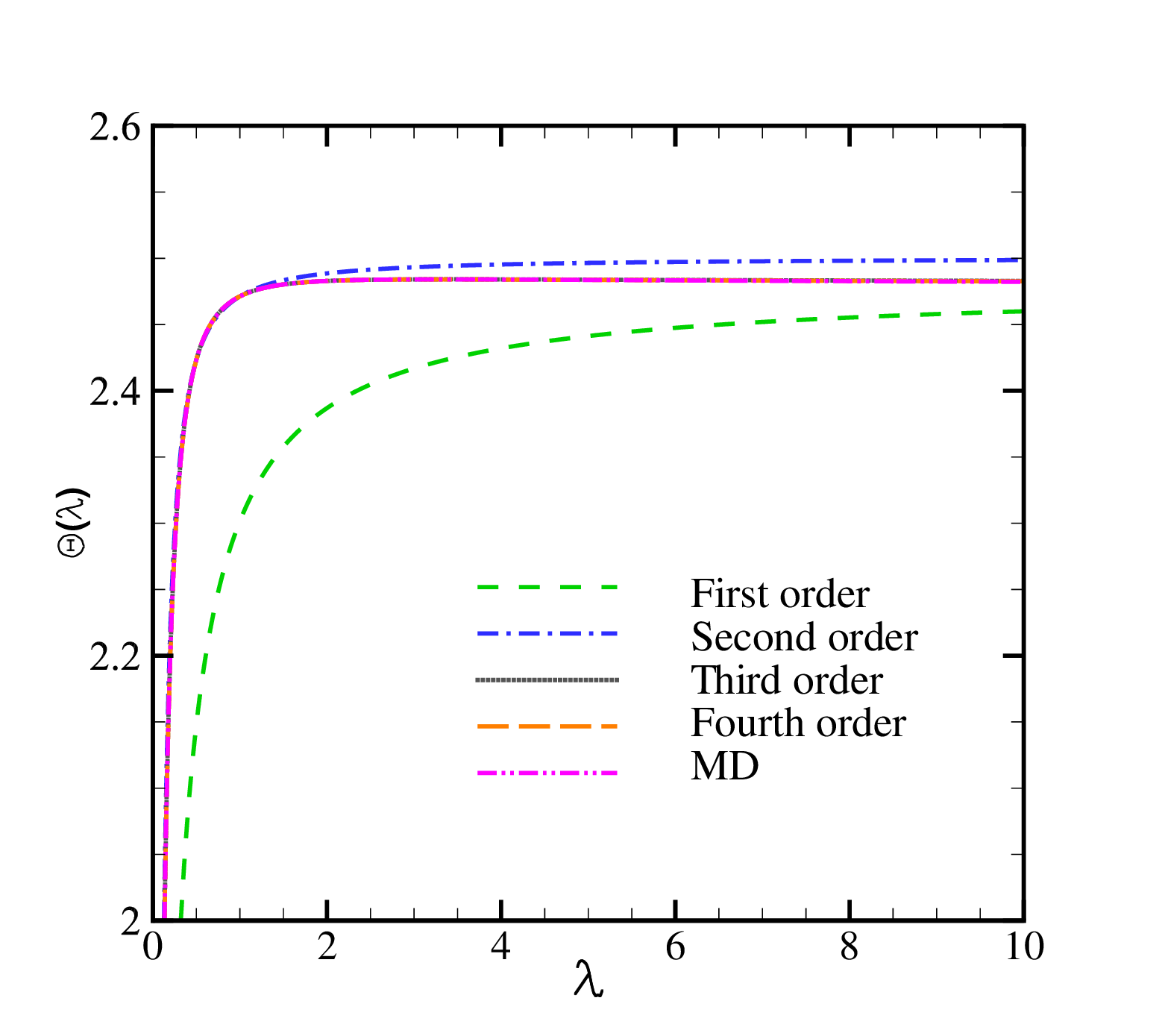}
	\includegraphics[scale=0.25]{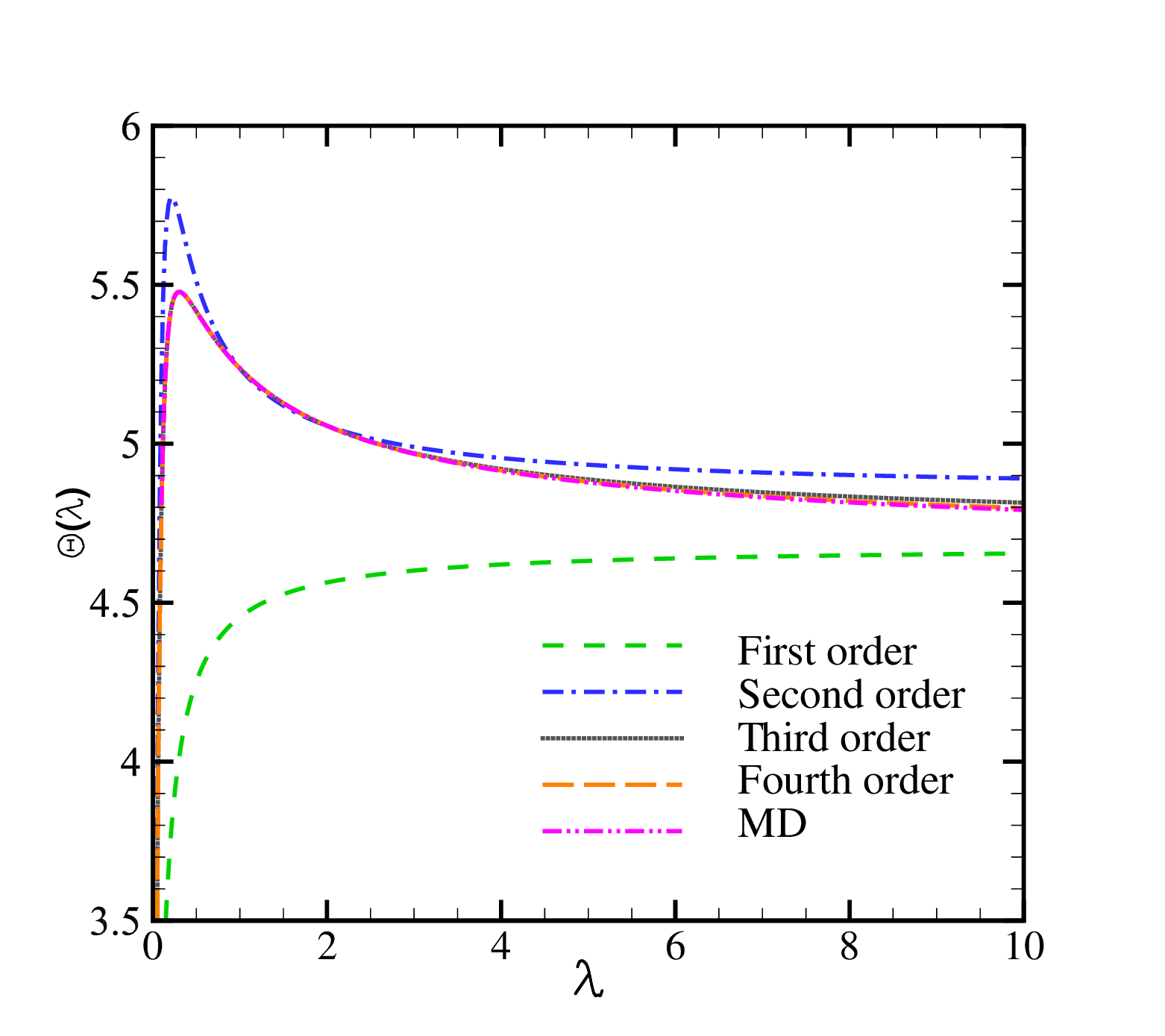}
	\caption{Laplace transform $\bm{\Theta}(\lambda)$ for a tagged particle in a particle bath obtained from full molecular dynamics data and calculated with different orders of rational approximation.
	Case (\Rmnum{1}) (left) and case (\Rmnum{2})(right).}
	\label{fig:tag_laplace_kernel}
\end{figure}

%\begin{equation}
%\begin{split}
%&\dot{\mathbf{q}} = \mathbf{p}/m, \\
%&\dot{\mathbf{p}} = -\beta \int_0^t \bm{\theta}(t-s)\mathbf{v}_q(s) ds +  \mathbf{R},
%\end{split}
%\label{eq:target_motion}
%\end{equation}
%where $\mathbf{R}$ is the fluctuation force on the target particle. We aim to compute the 
%memory term of $\bm{\theta}(t)$ using the velocity data.
%Theoretically, $\bm{\theta}(t)$ should be $3\times3$ matrix. However, since
%the system is homogeneous along $x$, $y$ and $z$ direction. $\bm{\theta}(t)$ reduces into
%a scalar. 
Following the Mori-Zwanzig method, the dynamics of the target particle can be modeled by a GLE with zero mean force.
We considered two cases: (\Rmnum{1}) $\beta = 0.5$, $a = 25.0$ and (\Rmnum{2}) $\beta = 1.0$, $a = 50.0$.
Based on the velocity correlation function $\langle \mb{v}(0)\mb{v}(t)^T\rangle$ obtained from MD simulation data, we can compute the different orders of kernel approximation $\bm{\Theta}(\lambda)$ by Eq.~\eqref{eq:Theta_second}.
As shown in Fig. \ref{fig:tag_laplace_kernel}, the exact kernel function $\bm{\Theta}(\lambda)$ agrees well with the numerical result directly obtained from $\langle \mb{v}(0)\mb{v}(t)^T \rangle$ as the approximation order $n$ increases to second order or above.

\begin{figure}[htbp]
	\center
	\includegraphics[scale=0.25]{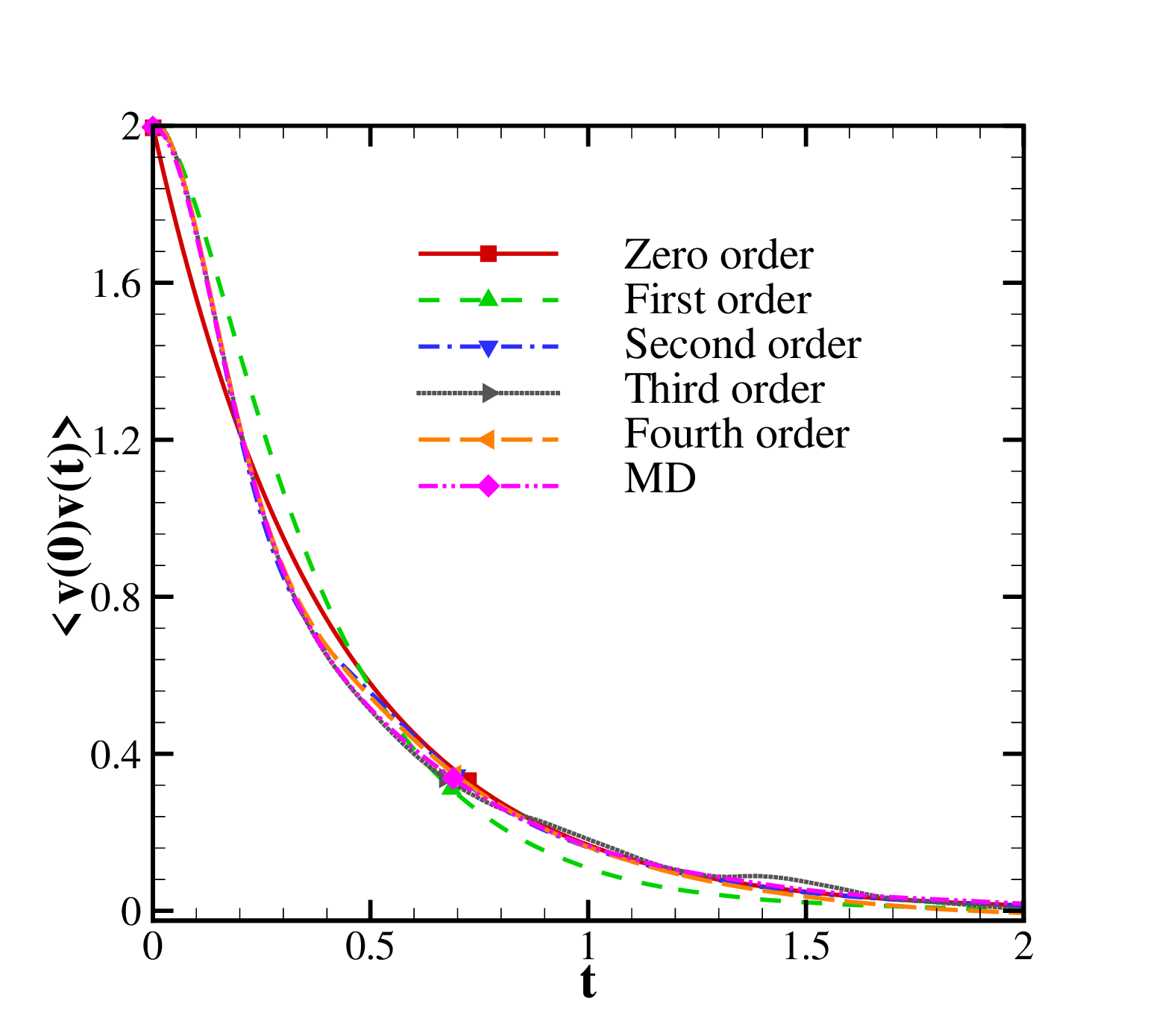}
	\includegraphics[scale=0.25]{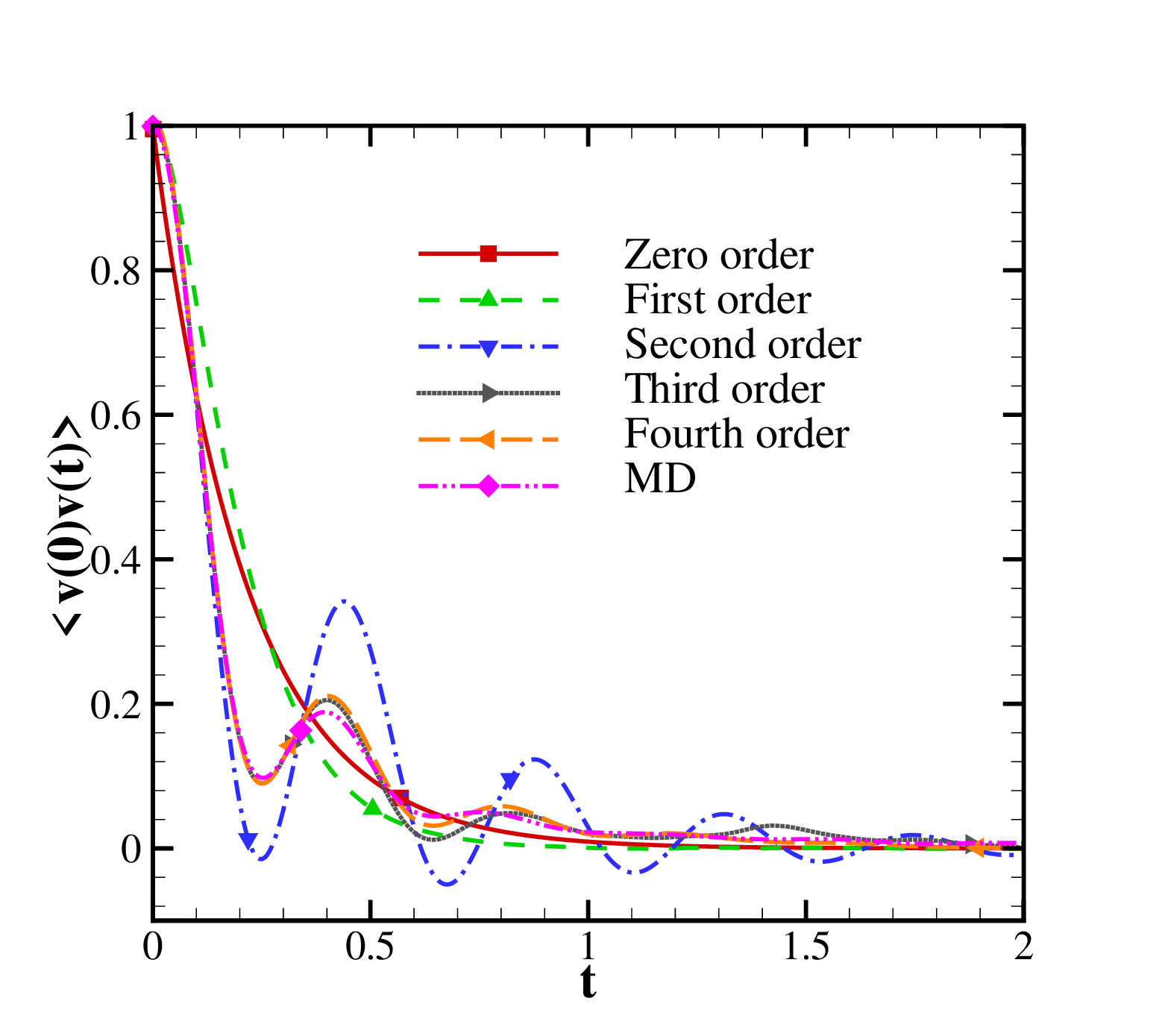}\\
	\includegraphics[scale=0.25]{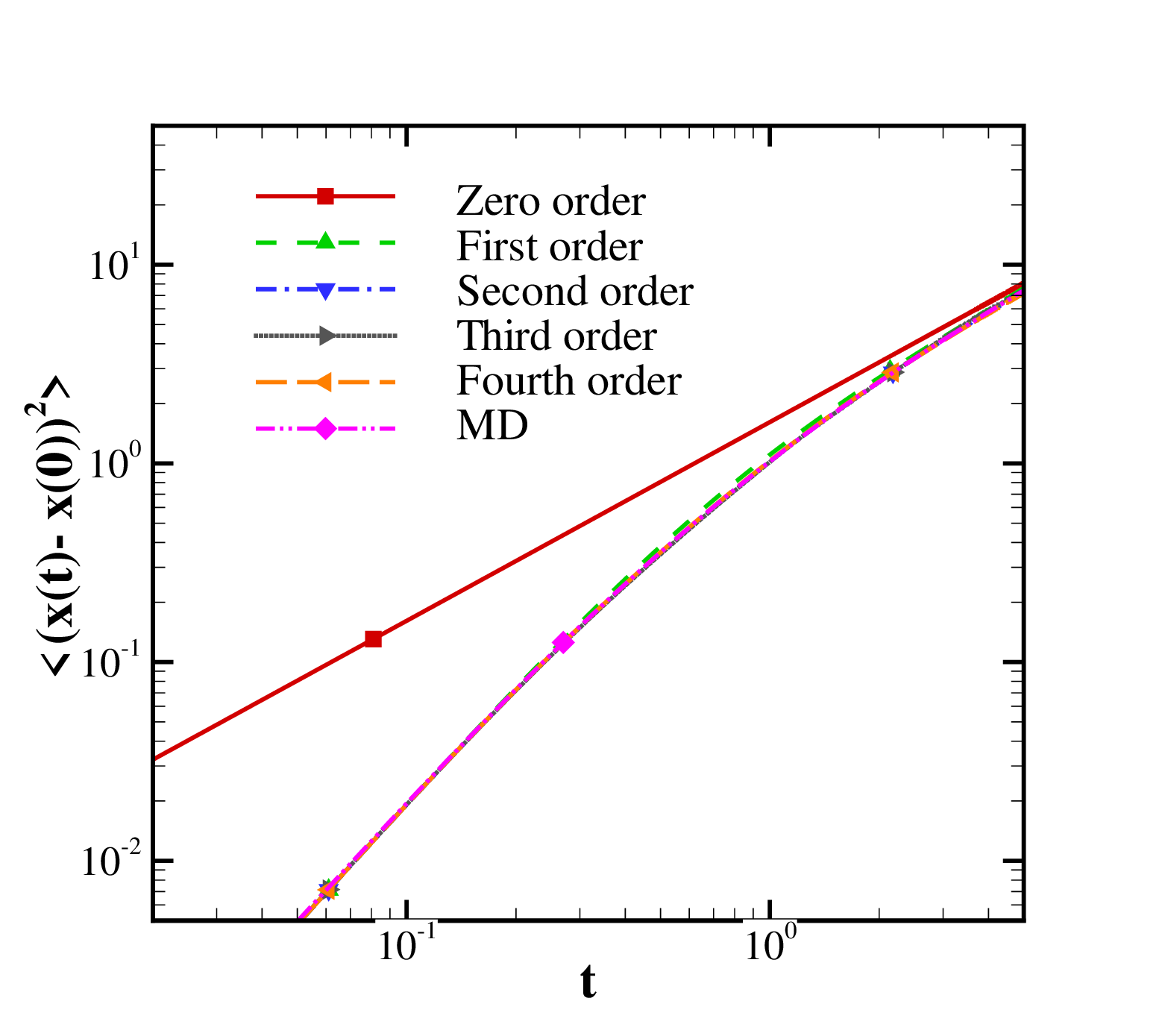}
	\includegraphics[scale=0.25]{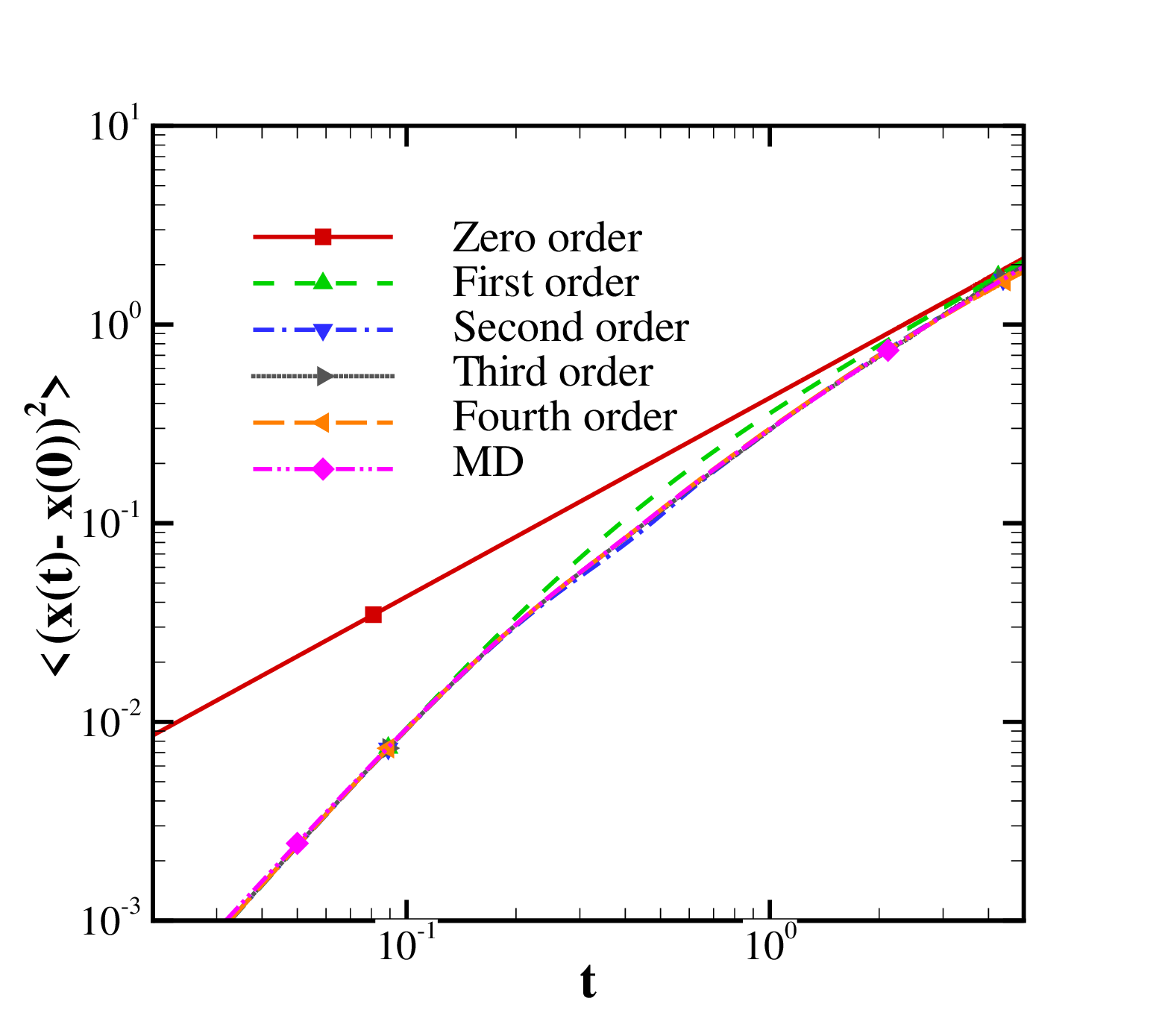}
	\caption{Simulations of a tagged particle in a particle bath.
	Upper: velocity correlation function obtained from MD data and different orders of the rational approximation for cases (\Rmnum{1}) (left) and  (\Rmnum{2}) (right).
	Lower: mean square displacement obtained from MD data and different order of the rational approximation for cases (\Rmnum{1}) (left) and (\Rmnum{2})(right).}
	\label{fig:tag_vcorr_msd}
\end{figure}

Unlike case (\Rmnum{1}), $\bm{\Theta}(\lambda)$ in case (\Rmnum{2}) shows a pronounced peak near $\lambda = 0.2$, indicating significant oscillations in the time domain of $\bm{\theta}(t)$.
Fig.~\ref{fig:tag_vcorr_msd} shows the velocity correlation function $\langle v(0)v(t)\rangle = \rm{Tr}\left[\langle \mb{v}(0)\mb{v}(t)^T\rangle\right]/3$ and the mean-squared displacement $\langle (q(t)-q(0))^2 \rangle = \rm{Tr}\left[\langle (\mb{q}(t)-\mb{q}(0)) (\mb{q}(t)-\mb{q}(0))^T \rangle\right]/3$ obtained from solving Eq.~\eqref{eq: gle} with kernel $\theta(t)$ constructed by different orders of rational approximations.
The zero-order approximation corresponds to the Markovian approximation of the kernel term
\begin{equation}
	\int_0^t \bm{\theta}(s) \mb{v}(t-s) ds \approx \left[\int_0^{\infty}\bm{\theta}(s)ds\right] \mb{v}(t).
	\label{eq:Markovian}
\end{equation}

\begin{figure}[htbp]
	\center
	\includegraphics[scale=0.36]{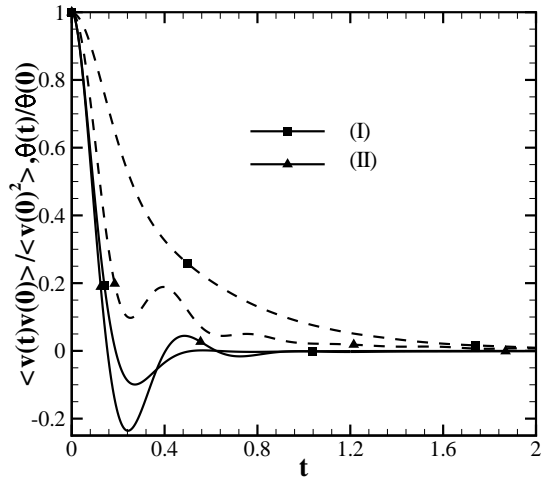}
	\caption{Simulations of a tagged particle in a particle bath.
	Velocity correlation (dashed line) and the fourth-order approximation of $\bm{\theta}(t)$ (solid line) for Cases (\Rmnum{1}) and (\Rmnum{2}).}
	\label{fig:tag_vcorr_kernel}
\end{figure}

In case (\Rmnum{1}), Eq. \eqref{eq: gle} can reproduce the dynamics of the system very well for rational approximations of second order (and above).
In contrast, for case (\Rmnum{2}), $\langle v(0)v(t)\rangle$ obtained from the second-order approximation exhibits artificial oscillation and deviations from the dynamics results; the third and fourth order approximations yield much better agreement.
The different performance for cases (\Rmnum{1}) and (\Rmnum{2}) can be understood by examining the time scale separation of $\bm{\theta}(t)$ and $\mb{v}(t)$ in the memory term $\displaystyle \int_0^t \bm{\theta}(s) \mb{v}(t-s) ds$.
As shown in Fig.~\ref{fig:tag_vcorr_kernel}, the velocity correlation $\langle v(0)v(t)\rangle$ of case (\Rmnum{1}) decays much more slowly than for case (\Rmnum{2}). The plateau region of $\bm{\Theta}(\lambda)$ of case(\Rmnum{1}) shown in Fig.~\ref{fig:tag_laplace_kernel} illustrates the similarities between $\bm{\theta}(t)$ and $\bm{\theta}_0\delta(t)$.
These similarities are why the Markovian approximation by Eq.~\eqref{eq:Markovian} yields fairly good agreement for case(\Rmnum{1}).
In contrast, there is no apparent time scale separation between $\mb{v}(t)$ and $\bm{\theta}(t)$ for case(\Rmnum{2}), explaining the need for higher-order approximations to characterize the coupling between $\bm{\theta}(t)$ and $\mb{v}(t)$.

\begin{figure}[htbp]
	\center
	\includegraphics[scale=0.36]{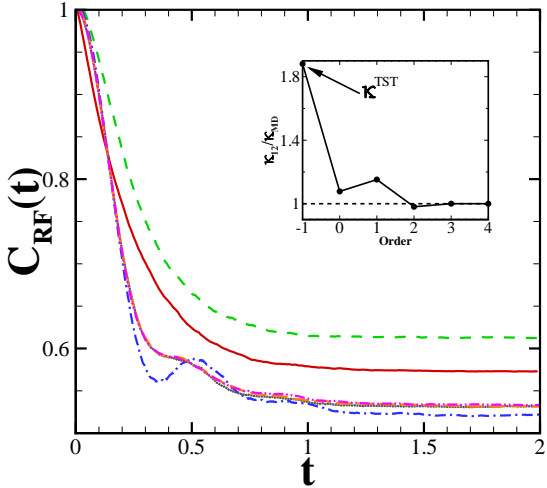}
	\caption{Transition flux correlations for a particle in a double-well potential obtained from full molecular dynamics and GLE simulations with different orders of approximation.
	The zero- to fourth-order approximations are represented by the solid, dash, dash-dot, dotted, long-dash lines.
	The full molecular dynamics result is represented by dash-dot-dot line.
	Inset: plateau reaction rate obtained by using GLEs with different orders of approximation.
	Order ``$-1$'' refers to the reaction rate predicted by transition state theory.}
	\label{fig:reaction}
\end{figure}

To further demonstrate our method, we simulated the transition rate of a tagged particle in a double-well potential using both by GLE and full molecular dynamics.
The double-well potential had the form
\begin{equation}
	U(x) = U_0\left[1-\left(\frac{x}{x_0}\right)^2\right]^2
\end{equation}
where $U_0 = 25$ and $x_0 = 2.5$ refer to the depth and width of the potential field.
The tagged particle interacted with solvent particles through interactions defined in Eq.~\eqref{eq:DPD_eq2} with $\beta = 1$
and $a = 50$, interactions which drove the transitions between the two energy minima at $x_1 = -x_0$ and $x_2 = x_0$.
The instantaneous transition rate $\kappa_{12}(t)$ was computed by 
\begin{equation}
	\kappa_{12}(t) = \langle \delta[S(0)-S_0]\dot{S}\chi(t)\rangle /Q_R,
\end{equation}
where $S$ is a collective variable defining the dividing iso-surface $S-S_0=0$ between the states and $\chi(t)$ is the characteristic function for the reaction trajectory.
For the double-well system, $S = x$ and $\chi(t) = H_s(S(t))H_s(-S(-t))$, where $H_s$ is the Heaviside function.
$Q_R = \displaystyle \int_{\Gamma_1} e^{-\beta H}$ is the reaction partition function over the phase space of state $1$. 
The transition flux correlation function $C_{RF}(t)$ is given by $C_{RF}(t) = \kappa_{12}(t)/\kappa^{\rm TST}$, where $\kappa^{\rm TST}$ is the reaction rate predicted from transition state theory.
Fig.~\ref{fig:reaction} shows $C_{RF}(t)$ obtained from the full molecular dynamics the GLE systems with kernels modeled by different orders of rational approximation.
The zero-order approximation (corresponding to Langevin dynamics with a constant friction coefficient) yielded a $C_{RF}(t)$ which deviated from the full molecular dynamics result, indicating a pronounced non-Markovian effect.
In contrast, $C_{RF}(t)$ obtained from the GLE using the third- and fourth-order rational approximations agree well with the full molecular dynamics results.
Moreover, as shown in the inset of Fig.~\ref{fig:reaction}, both transition state theory and the zero-order (Langevin dynamics) model overestimate the reaction rate; $\kappa_{12}$ approaches the full molecular dynamics results only when using the third- and fourth-order approximations where the memory kernel can be more accurately modeled.

\section{Summary}
We have presented a new data-driven approach to obtain the memory kernel for a GLE through rational-function approximation in the Laplace transform domain.
The data-driven nature of this method arises through connection of the rational function coefficients with equilibrium statistics of the coarse-grained variables -- statistics that can be calculated through simulation time series data.
The zero- and first-order approximations recover the Langevin and Ornstein-Uhlenbeck stochastic processes, respectively.
Higher-order approximations have also been systematically derived for systems with significant memory effects.
Unlike the time-domain kernel function representation, numerical simulations of the GLE using the rational approximation can be conveniently implemented by introducing auxiliary variables that follow linear stochastic dynamics with no memory, eliminating the need for expensive calculations of history-dependent memory terms.
We have also shown that the second FDT \eqref{eq: fdt} can be satisfied automatically using our approach.
This method has been tested with simple systems but applicable to much more complicated biological and material systems with pronounced memory effects; e.g., sub-diffusion in single-molecule measurements \cite{Kou_Xie_PRL_2004, Kou_Xie_PRL_2005} or transition dynamics of chemical and biological reaction systems \cite{Schenter_JCP_1992, lange2006collective}.
The rational function approximations (Eq. \eqref{eq: gle1} and Eq. \eqref{eq: gle2}) of the GLE enable us to analyze the transition dynamics via the extended dynamics in an augmented  phase space $(\mathbf{q}, \mathbf{p}, \mathbf{d})$, where direct analysis via the GLE could be difficult/inaccessible.

\begin{acknowledgments}
We thank George Karniadakis, Eric Darve, Panos Stinis, Greg Schenter, Lei Wu, Dave 
Sept, J.~Andrew McCammon, and Zhen Li for helpful discussion. This work was supported 
by the U.S. Department of Energy, Office of Science, Office of Advanced Scientific Computing 
Research as part of the Collaboratory on Mathematics for Mesoscopic 
Modeling of Materials (CM4).
\end{acknowledgments}

\bibliographystyle{apsrev4-1}
\bibliography{main}

%\appendix
%
%\textcolor{red}{
%Fig. \ref{fig:tag_vcorr_kernel} shows the velocity 
%correlation $\langle \mb{v}(0)\mb{v}(t)\rangle$ and $\bm{\theta}(t)$ (the fourth order approximation) for both cases.
%For case(\Rmnum{1}), $\langle \mb{v}(0)\mb{v}(t)\rangle$ decays relatively slower than $\bm{\theta}(t)$. Therefore, Markovian 
%approximation by Eq. \eqref{eq:Markovian} yields fairly good approximation of the memory term. However, for 
%case(\Rmnum{2}), the time scale of $\langle \mb{v}(0)\mb{v}(t)\rangle$ is similar to $\bm{\theta}(t)$. Therefore, high order
%approximation is required for characterize the coupling of $\bm{\theta}(t)$ and $\mb{v}(t)$. 
%}
%%Finally, we note that the large ``dimple''
%%region of $\bm{\theta}(t)$ for case(\Rmnum{2}) also explains the larger peak region of $\bm{\Theta}(\lambda)$ near $\lambda = 0.2$
%%in Fig. \ref{fig:tag_laplace_kernel}.
%
%\begin{figure}[htbp]
%\center
%\includegraphics*[scale=0.24]{./Figure/Group_3/Vcorr_Kernel_compare_A_50_A_25.eps}
%\caption{Velocity correlation (dash line) and the fourth order approximation of $\bm{\theta}(t)$ (solid line) 
%  for case(\Rmnum{1}) and case(\Rmnum{2}).}
%\label{fig:tag_vcorr_kernel}
%\end{figure}

\end{document}